\documentclass[preprint,5p,twocolumn]{elsarticle}
\pdfoutput=1
\usepackage{geometry}
\usepackage{graphicx}
\usepackage{hyperref}
\usepackage{natbib}
\usepackage{bm}
\begin{document}
\begin{frontmatter}
\title{Real-Time Detection of Gravitational Waves from Binary Neutron Stars using
Artificial Neural Networks}
\author{Plamen G. Krastev}
\ead{plamenlrastev@fas.harvard.edu}
\address{Harvard University, Faculty of Arts and Sciences, Research Computing, \\ 38 Oxford Street, Cambridge, MA 02138, U.S.A.}
\begin{abstract}
The groundbreaking discoveries of gravitational waves from binary black-hole mergers \cite{Abbott:2016blz,Abbott:2016nmj,Abbott:2017vtc}
and, most recently, coalescing neutron stars \cite{TheLIGOScientific:2017qsa} started a new era of Multi-Messenger Astrophysics
and revolutionized our understanding of the Cosmos. Machine learning techniques such as artificial neural networks are
already transforming many technological fields and have also proven successful in gravitational-wave astrophysics for detection and
characterization of gravitational-wave signals from binary black holes \cite{George:2016hay,George:2017pmj,Gabbard:2017lja}. Here we use 
a deep-learning approach to rapidly identify transient gravitational-wave signals from binary neutron star mergers in noisy time series
representative of typical gravitational-wave detector data. Specifically, we show that a deep convolution neural network trained on
100,000 data samples can promptly identify binary neutron star gravitational-wave signals and distinguish them from noise and signals
from merging black hole binaries. These results demonstrate the potential of artificial neural networks for real-time detection
of gravitational-wave signals from binary neutron star mergers, which is critical for a prompt follow-up and detailed observation
of the electromagnetic and astro-particle counterparts accompanying these important transients.
\end{abstract}
\end{frontmatter}

\section{Introduction}

The detections of gravitational waves (GWs) from binary black hole (BBH) mergers have verified Einstein's theory of General Relativity
in extraordinary detail in the most violent astrophysical environments \cite{Abbott:2016blz,Abbott:2016nmj,Abbott:2017vtc,Abbott:2017gyy}. 
In addition, the first observation of coalescing neutron stars in both gravitational and electromagnetic spectra has initiated 
the era of Multi-Messenger Astrophysics (MMA), which uses observations in electromagnetic radiation, gravitational 
waves, cosmic rays, and neutrinos to provide deeper insights about properties of astrophysical objects and phenomena 
\cite{TheLIGOScientific:2017qsa,GBM:2017lvd}. These discoveries were made possible by the Advanced Laser 
Interferometer Gravitational Wave Observatory (LIGO) and Virgo collaborations. As gravitational-wave detectors increase 
their sensitivity many more observations, including BBH, binary neutron star (BNS) and black hole - neutron star
(BHNS) signals are likely to be detected more frequently. Conventional gravitational-wave detection techniques are based mainly
on a method known as template matched filtering \cite{Gabbard:2017lja,Canton:2014ena}, which typically uses large banks of 
template waveforms each with different compact binary parameters, such as component masses and/or spins. Since parameters 
are not known in advance, a template bank spans a large astronomical parameter space, which makes these approaches very 
computationally expensive and challenging. In particular, as it has been already pointed out in the literature 
\cite{Harry:2016ijz,Huerta:2016rwp}, GW searches based on matched-filtering techniques currently target a 4D parameter 
space (compact binary sources with spin-aligned components on quasi-circular orbits) out of the 9D parameter space 
available to the current GW detectors (binary component masses ($m_1$, $m_2$) and spins ($\bm{\hat{s}_1}$, $\bm{\hat{s}_2}$) 
plus the orbital eccentricity $e$). The computational cost of these low-latency GW searches based on implementations 
of matched-filtering is presently such that their extension to the full 9D signal manifold is computationally 
prohibitive \cite{Huerta:2019rtg}. Most importantly, these surveys may miss important GW transients where a rapid
follow-up is critical for successful observation of their electromagnetic counterparts. Specifically, the optical counterparts of 
gravitational waves from the merger of BNS and BHNS systems, known as kilonovae \cite{Metzger:2011bv}, encode key information 
required to constrain the physical properties of the transient, but due to their fast decay rate they need to be identified and 
localized within several hours after the compact binary merger and promptly observed in the entire electromagnetic spectrum. 
Therefore, based on the above considerations, the need arises for new methods to overcome the limitations and computational challenges
of existing GW detection algorithms, in particular, approaches to detect in real-time GW signals from binary neutron star (and
black hole-neutron star) mergers in the full parameter space available to current and future GW detectors. 

In this work, we explore a deep-learning approach to rapidly detect gravitational - wave signals from binary neutron star mergers.
Deep learning algorithms \cite{DL}, a subset of machine learning, have been very successful in tasks, such as image recognition
\cite{DL,DLVis} and natural language processing \cite{DL_NLP}, and recently also emerged as a new tool in GW astrophysics for 
detection, characterization \cite{George:2016hay,George:2017pmj,Gabbard:2017lja} and denoising \cite{Wei:2019zlc} of GW signals 
from binary black holes. Deep-learning methods are able to perform analysis rapidly since the computationally intensive part of 
the algorithm is done during the training stage before the actual data analysis \cite{DL_Book}, which could make them orders 
of magnitude faster than conventional match-filtering techniques \cite{George:2016hay}. Here, we demonstrate the power of 
the deep-learning approach on the specific example of rapidly classifying gravitational waves from binary neutron star mergers 
from detector noise and signals from binary black holes. This example shows clearly that machine learning can help in the real-time 
detection of BNS signals and thus trigger a prompt follow-up of the electromagnetic counterparts of the gravitational-wave 
transient.

\section{Methods}

Deep learning algorithms consist of processing units, neurons, which are arranged in arrays forming one to several layers.
A neuron acts as a filter performing a linear operation between the input array and the weights associated with the neuron.
A deep neural network has an input layer, typically followed by one or more hidden layers, and a final layer with one or more
output neurons. In classification problems, the output neurons give the probabilities that an input sample belongs to a
specific class. In this case, we distinguish between three classes of time series, BNS and BBH merger signals in additive
Gaussian noise (signals plus noise), and Gaussian noise only, where we use integer class labels (O: Noise, 1: BBH signal, 
2: BNS signal). Accordingly, the data sets consist of simulated gravitational-wave time series where the compact 
binary merger signals (BNS and BBH) are generated using the LIGO Algorithm Library--LALSuite \cite{LALSuite}. 
For the BNS signals, we use the PhenomPNRT waveform model \cite{Abbott:2018wiz} and simulate systems with 
component masses in the range from 1 to $2M_{\odot}$, including also tidal deformation contributions, where the 
tidal deformability, $\Lambda$, is computed with the APR equation of state (EOS) \cite{Akmal:1998cf}. 
(For computing $\Lambda$, see e.g,. Refs. \cite{Hinderer:2009ca,Krastev:2018wtx}.) The BBH signals are simulated 
using the SEOBNRv2 waveform model \cite{Purrer:2015tud}, which models the inspiral, merger and ringdown components 
of the signal. We simulate systems with component masses in the range from 5 to $50M_{\odot}$, with zero spin. 
The simulated signals are chosen to be 10 seconds in duration sampled at 4096 Hz. This choice was made because 
BNS signals are considerably longer and contain typically much higher frequencies than BBH gravitational-wave signals.

\begin{figure}[t!]
\centering
\includegraphics[scale=0.5]{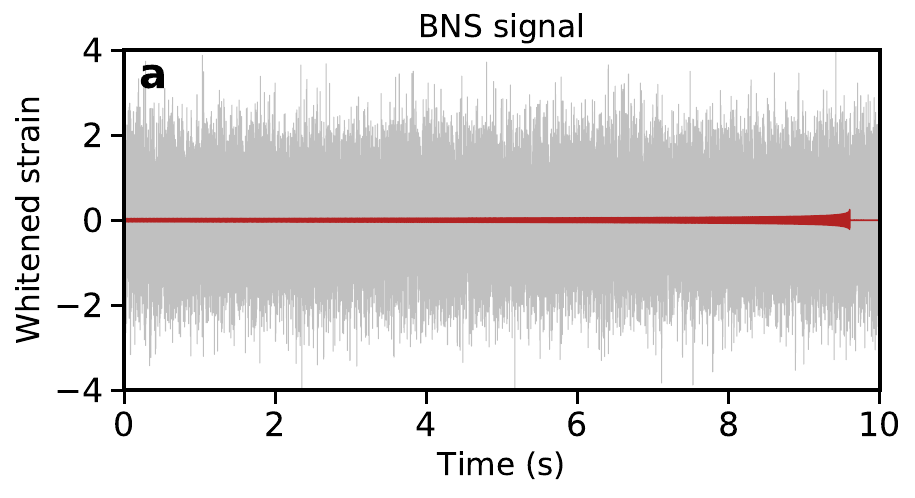}
\includegraphics[scale=0.5]{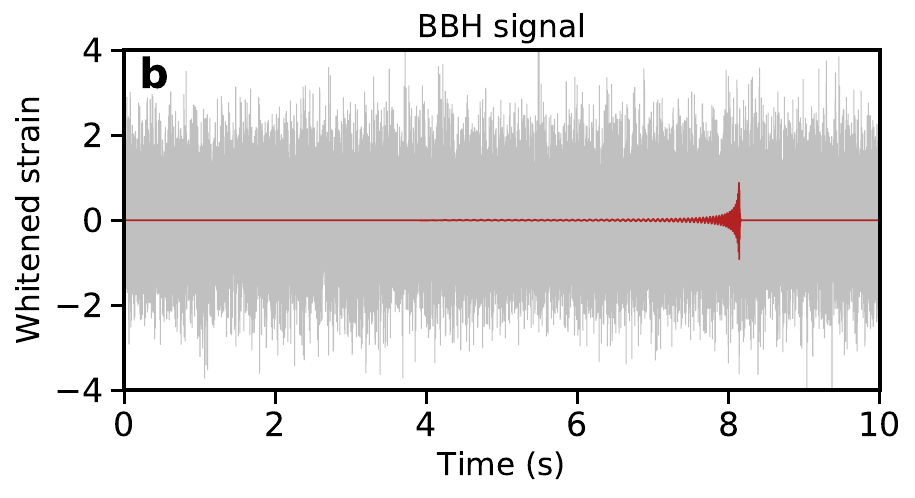}
\caption{{\bf Sample BNS and BBH signals injected in simulated Gaussian noise.} {\bf (a)} A whitened noise-free time 
series of a binary neutron star gravitational-wave signal with component masses $m_1=1.4M_{\odot}$ and $m_2=1.6M_{\odot}$ 
and dimensionless tidal deformability  $\Lambda_1=261.9$ and $\Lambda_2=105.5$ (computed with the APR equation of state 
(EOS) \cite{Akmal:1998cf}) with optimal SNR $\rho_{opt} = 8$ (dark red). The gray time series shows the same 
gravitational-wave signal with additive white Gaussian noise of unit variance. This time series is an example of the 
data sets used to train, validate and test the convolutional neural network. {\bf (b)} Same as (a) but for a binary 
black-hole gravitational-wave signal with component masses $m_1=27M_{\odot}$ and $m_2=49M_{\odot}$. 
($\Lambda=0$ for black holes.)}\label{fig1}
\end{figure}

The simulated signals are "whitened" with Advanced LIGO's power spectral density (PSD) at the "zero-detuned high-power"  
\cite{LIGO-LRR-2016} to rescale the noise contribution at each frequency to have equal power \cite{Gabbard:2017lja}. 
Subsequently, the waveforms are shifted randomly such that the peak amplitude of each waveform is randomly positioned 
in the range from 9.65 to 9.95 seconds of the time series for the BNS signals, and from 8 to 9.95 seconds for the 
BBH signals (since BBH signals are considerably shorter than BNS signals), to reassure robustness of the network 
against temporal translations. Different realizations of white Gaussian noise are superimposed on top of the signals, 
while the waveform amplitude is scaled to achieve a predefined optimal signal-to-noise ratio (SNR) defined as \cite{Gabbard:2017lja}
\begin{equation}
\rho_{opt}^2 = 4\int_{f_{min}}^{\infty}\frac{|\tilde{h}(f)|^2}{S_n(f)}df, 
\end{equation}
where $\tilde{h}(f)$ is the frequency domain representation of the GW strain, $S_n(f)$ is the single-sided detector noise
PSD (chosen here at the "zero-detuned high-power") and $f_{min}$ is the frequency of the GW signal at the start of
the sample time series. From an astrophysical perspective, rescaling the GW waveform simply translates to moving
the source closer or further away from the detector. Example time series are shown in Fig. \ref{fig1}.

\begin{figure*}[t!]
\centering
\includegraphics[scale=0.45]{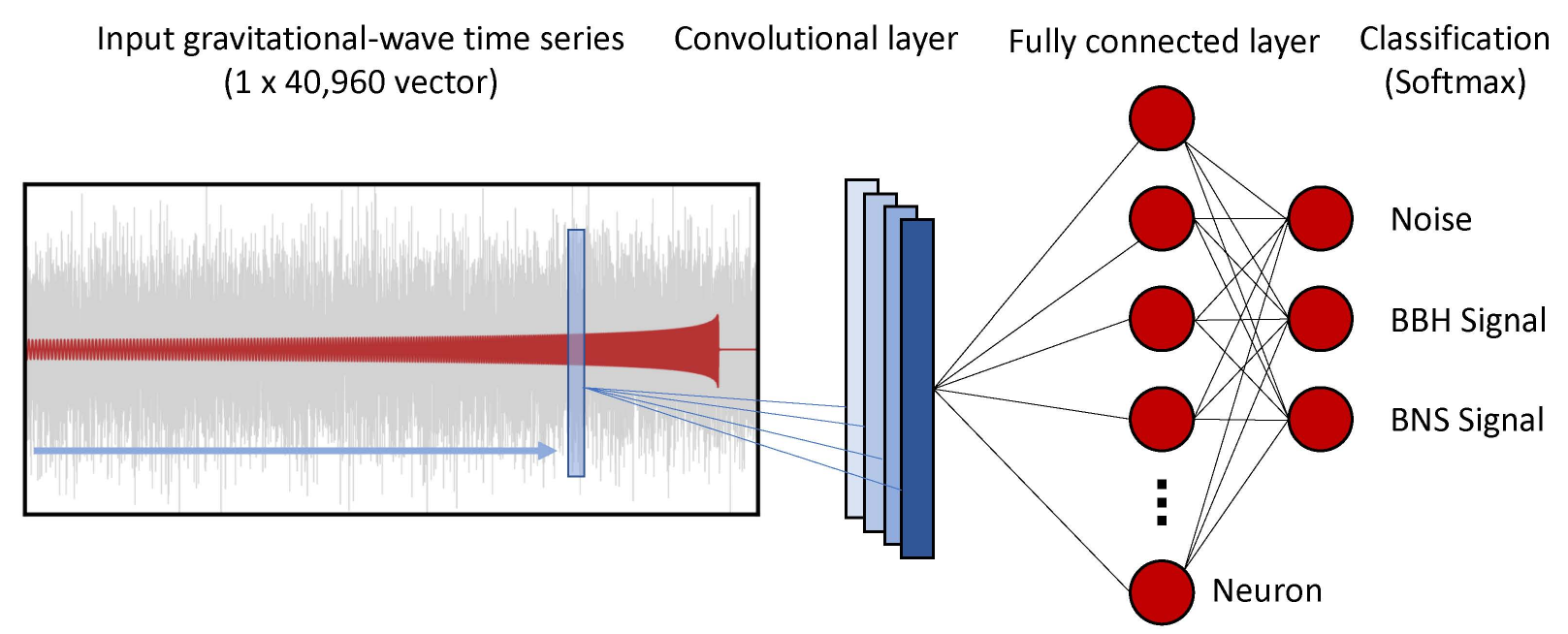}
\caption{{\bf Using an artificial neural network to detect gravitational-wave signals from binary neutron-star mergers.}
Gravitational-wave time series serve as input for a deep convolutional network with convolutional and fully connected layers.
The blue arrow represents the sliding of the convolutional filters along the input time-series vector. The last softmax layer
outputs the probability that the input time series belongs to a certain class (Noise, BBH Signal, or BNS Signal).
The weights of the artificial neural network are tuned by training on many labeled data samples and the network can then
classify an unknown sample time series with high confidence.}\label{fig2}
\end{figure*}

Supervised learning requires that data sets are divided into training, validation and testing data. Training data is
used by the network to learn from, validation data allows for verification of whether the network is learning correctly,
and the testing data is used to assess the performance of the trained model. The training sets used here consist of
100,000 independent time series with 1/3 containing BNS signal + noise, 1/3 BBH signal + noise, and 1/3 noise only. The
validation and testing data sets each consist of 5,000 independent samples containing (approximately) equal fractions of each
time-series class. To ensure that the neural network can identify BNS gravitational-wave signals over a broad range of
astronomically motivated SNR values, we start the network training with large SNR and then gradually reduce the
SNR to lower levels. This approach is adapted from "curriculum learning" \cite{Shen:2019vep} and enables the network to
learn to distinguish signals with lower SNR more accurately. Specifically, for all training sessions the SNR of each BBH and BNS 
waveform was randomly sampled in the range [SNR$_{low}$, SNR$_{high}$] with SNR$_{high}$ = 20. Initially, SNR$_{low}$ was 
set to 20 and then gradually decreased to 3 in steps of 1 in each subsequent training session. Thus, the final SNR 
was uniformly sampled in the range between 3 and 20. 

The neural network used here is a convolutional neural network (CNN) \cite{CNN} and has 4 convolutional and 4 pooling layers,
followed by 2 dense fully connected layers. The filter sizes of the convolutional layers are 32, 64, 128 and 256
respectively, and the sizes of the dense layers are 128 and 64. We used kernel sizes of 16, 8, 8 and 8 for the 
convolutional layers and 4 for all pooling layers. The first layer corresponds to the input to the neural network 
which in this case is a one-dimensional time-series vector (of dimension 40,960). Each neuron in the convolutional layers 
computes the convolution between the neuron's weight vector and the outputs from the layer below. Neuron weights are updated
through an optimization back propagation algorithm \cite{BP}. Pooling layers perform a down-sampling operation along
the spatial dimensions of their input. At the end, there is a hidden dense layer connected to an output layer computing the
inferred class probabilities. The network design is optimized by fine-tuning multiple hyper-parameters, which include
here the number and type of network layers, the number of neurons in each layer, max-pooling parameters, and type of
activation functions. The optimal network architecture was determined through multiple experiments and tuning of the
hyper-parameters. We show the architecture of the final network in Table~\ref{tab1}. 

To build and train the neural network, the Python toolkit Keras (https://www.tensorflow.org/guide/keras) was used, which 
provides a high-level programming interface to access the TensorFlow \cite{TF} (https://www.tensorflow.org) deep-learning 
library. We use the technique of stochastic gradient descent with an adaptive learning rate with the ADAM method \cite{ADAM}
with the AMSgrad modification \cite{ADAM2}. To train the neural network, we use an initial learning rate of 0.001 and choose 
a batch size of 1000. For each SNR range the number of training epochs is limited to 100, or until the error on the validation 
data set stops decreasing. The network training was performed on NVIDIA Tesla V100 GPU and the size of the mini-batches 
was chosen automatically depending on the specifics of the GPU and data sets. The cost function was selected to be the 
sparse categorical cross-entropy loss. The process of using an artificial neural network to detect gravitational-wave signals 
from BNS mergers is illustrated in Fig.~\ref{fig2}.

\begin{table}[t!]
\begin{center}
\begin{tabular}{r@{\hskip 3mm}lc@{\hskip 3mm}l}
    & Layer            & Array Type & Size  \\
    \hline
    & Input            & Vector     & 40960      \\
 1  & Reshape          & Matrix     & 1 x 40960  \\
 2  & Convolution (1D) & Matrix     & 32 x 40945 \\
 3  & Pooling          & Matrix     & 32 x 10236 \\
 4  & ReLU             & Matrix     & 32 x 10236 \\
 5  & Convolution (1D) & Matrix     & 64 x 10229 \\
 6  & Pooling          & Matrix     & 64 x 2557  \\
 7  & ReLU             & Matrix     & 64 x 2557  \\
 8  & Convolution (1D) & Matrix     & 128 x 2550 \\
 9  & Pooling          & Matrix     & 128 x 637  \\
 10 & ReLU             & Matrix     & 128 x 637  \\
 11 & Convolution (1D) & Matrix     & 256 x 623  \\
 12 & Pooling          & Matrix     & 256 x 155  \\
 13 & ReLU             & Matrix     & 256 x 155  \\
 14 & Flatten          & Vector     & 39680      \\
 15 & Dense Layer      & Vector     & 128        \\
 16 & ReLU             & Vector     & 128        \\
 17 & Dense Layer      & Vector     & 64         \\
 18 & ReLU             & Vector     & 64         \\   
 19 & Dense Layer      & Vector     & 3          \\
    & Output (Softmax) & Vector     & 3          \\
\hline
\end{tabular}
\end{center}
\caption{{\bf Architecture of the deep neural network.} The architecture of the deep one-dimensional convolutional
network consist of input layer, followed by 19 hidden layers, and output {\it softmax} layer. The size of the network
is about 83 MB.}\label{tab1}
\end{table}

\section{Results}

We assess the performance of the neural network by extracting the probability values produced by the neurons in the
output layer. These values are between 0 and 1 with their sum being unity. Each neuron gives the inferred probability
that the input time series belong to the noise, BBH signal, or BNS signal class, respectively. Specifically, for a given
SNR we construct the receiver operator characteristic (ROC) curves for the BBH and BNS classes. Here a ROC curve 
represents the fraction of signals correctly identified as their respective class, BNS or BBH (true alarm probability), 
versus the fraction of samples identified incorrectly as signals of the particular class (false alarm probability). A ranking 
statistic is considered superior to another if at a fixed false alarm probability it reaches a higher true alarm probability 
(sensitivity) \cite{Gabbard:2017lja}. The ROC curves are conveniently constructed with the Python scikit-learn library 
(https://scikit-learn.org). The optimal SNR was varied from 1 to 20 in integer steps of 1 and the classifier was applied to 
time-series inputs containing approximately equal fractions of each class (Noise, BBH Signal, BNS Signal).
 
\begin{figure}[t!]
\centering
\includegraphics[scale=1.0]{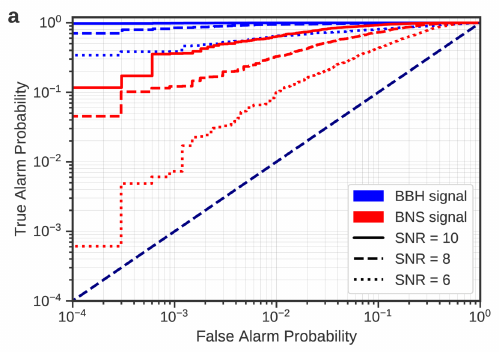}
\includegraphics[scale=1.0]{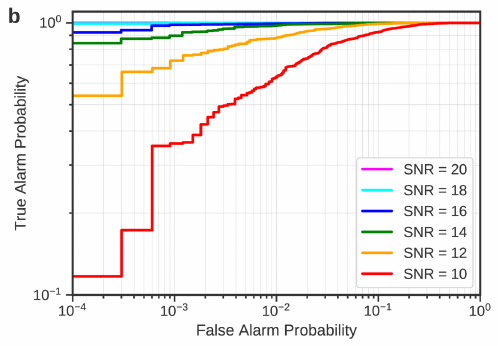}
\caption{{\bf  ROC curves.}  {\bf (a)}  ROC curves for test data sets containing BBH and BNS GW signals with
optimal SNR, $\rho_{opt} = 6, 8, 10$. The true alarm probability is shown versus the false alarm probability 
estimated from the output of the convolutional neural network. {\bf (b)} Same as (a) but only for BNS GW signals 
with  optimal SNR, $\rho_{opt}$,  in the range from 10 to 20 in steps of 2.}\label{fig3}
\end{figure}

In Fig. \ref{fig3} we show the ROC curves calculated for test data sets containing BBH and BNS GW signals. These results
show that the neural network is more sensitive to detecting GW signals from BBH than BNS mergers. In particular,
the neural network reaches a maximal true alarm probability for BBH signals with optimal SNR $\rho_{opt} = 10$ 
across the range of false alarm probabilities explored in this work (Fig. \ref{fig3} (a)). On the other hand, it achieves
a maximal true alarm probability for BNS signals with optimal SNR $\rho_{opt} = 18$ (Fig. \ref{fig3} (b)). The results imply 
that all BBH signals are identified for SNR $\geq$ 10 and both BBH and BNS signals are detected for SNR $\geq$ 18. 

\begin{figure}[t!]
\centering
\includegraphics[scale=1.0]{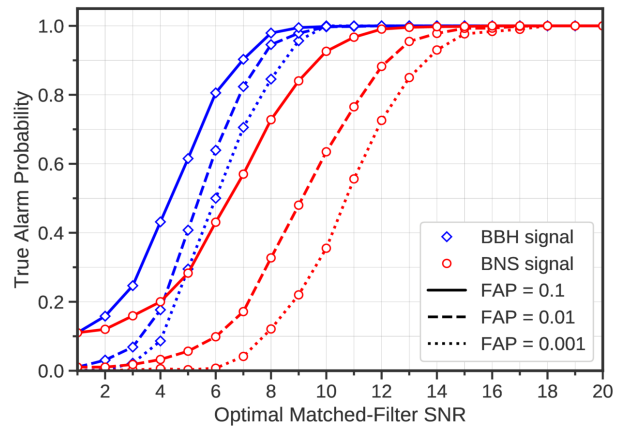}
\caption{{\bf Sensitivity curves illustrating the ability of the neural network to identify BNS and BBH GW signals.}
The true alarm probability is plotted as a function of the optimal SNR for false alarm probabilities $10^{-1}$, $10^{-2}$, 
and $10^{-3}$.}\label{fig4}
\end{figure}

We analyze the performance of the classifier further by looking at the sensitivity of detection of BNS and BBH signals 
for different SNR values at a fixed false alarm probability. These sensitivity curves are shown in Fig. \ref{fig4}, where 
the true alarm probability (sensitivity) is plotted as a function of the optimal SNR for several false alarm probabilities ($10^{-1}$, 
$10^{-2}$, $10^{-3}$). The sensitivity of detection of the neural network to identify GW signals from BBH mergers is
very similar to the one reported by  Gabbard {\it et al.} \cite{Gabbard:2017lja}. It is also seen in Fig. \ref{fig4} 
that all curves saturate (at 1) for optimal SNR $\geq$ 18, i.e., all signals, both BNS and BBH, are always detected.

Furthermore, the deep neural network automatically extracts and compresses information by finding patterns in the training 
data, dramatically reducing data dimensionality and thus creating a very computationally efficient and portable model. For instance, 
the size of the trained model is only about 83 MB, including the network weights and architecture information, therefore 
compressing approximately $6.6\times 10^4$ gravitational BBH and BNS waveforms (excluding noise samples), each of duration 
10 seconds sampled at 4096 Hz, with a total size of about 10.4 GB. The trained neural network model can therefore be viewed
as an abstract and compact representation of the template bank. In addition, the computational cost of evaluating the neural
network on new GW data, after it has been trained, does not depend on the data set size. As mentioned
previously, the computationally intensive training stage is performed only once offline. For example, once trained, 
processing 10 seconds of gravitational-wave data takes only milliseconds on both CPUs and GPUs with the final
CNN architecture. Such rapid processing is advantageous for generating real-time alerts and can provide useful
hints for follow up searches of electromagnetic counterparts of GWs and also for focused analysis with accurate
matched filtering approaches and Bayesian parameter estimation \cite{Gebhard:2019ldz}. For instance,
as more GW detectors come online, the computational cost of matched filtering methods scales at least
linearly in the number of detectors. (This is because the search for triggers is first performed independently for
each detector.) Moreover, the computational cost for trigger generation also scales linearly in the number of waveforms
in the template banks. As template banks become bigger, matched filtering becomes increasingly computationally
expensive, which makes online real-time trigger generation very computationally challenging \cite{Gebhard:2019ldz}. 
Specifically, the extension of real-time matched filtering techniques to the full 9D signal manifold currently available to 
GW detectors is computationally prohibitive \cite{Huerta:2019rtg}. These computational considerations are the major 
motivation to explore alternative detection methods in the first place. 

In addition, a real-time detection of GWs from compact binary systems involving  neutron stars would enable fast source 
localization necessary for rapid multi-wavelength follow-ups with relevant telescopes, which have small fields of view. 
The results of this study demonstrate the potential of deep learning algorithms to aid the prompt detection of GW signals 
from binary neutron star mergers and distinguish them from BBH signals and noise over a wide SNR range, with moderate 
computing resources (e.g., a standard laptop computer), which would make it possible to trigger timely and detailed 
observations of their electromagnetic counterparts.   

\section{Summary}

In conclusion, we have demonstrated the detection of gravitational waves from binary neutron star mergers via
deep learning techniques on simulated gravitational-wave detector data using the specific example of data
containing BNS and BBH signals in Gaussian noise. These results point the way to real-time detection of
gravitational waves from multi-messenger astrophysical sources, where a rapid follow-up is critical. Future
directions include using machine learning algorithms for real-time parameter estimation of gravitational-wave
signals from BNS and BHNS systems. In particular, machine learning approaches could help to extract challenging
source parameters, e.g., neutron-star tidal deformability \cite{Hinderer:2009ca}, which is extremely important for
understanding properties of dense matter and fundamental interactions, but is theoretically controversial and
observationally challenging to deduce.

\section*{Acknowledgements}
I thank Brendan Meade for stimulating my interest in machine learning and for useful discussions. I also thank Edo Berger for
valuable conversations on topics discussed in this work. In addition, I would like to thank the anonymous referee for 
carefully reading the manuscript and providing important feedback. This research has made use of software tools obtained from 
the Gravitational Wave Open Science Center (https://www.gw-openscience.org). The computational resources were provided by the 
Faculty of Arts and Sciences Research Computing at Harvard University.


\begin{thebibliography}{100}

%\cite{Abbott:2016blz}
\bibitem{Abbott:2016blz}
  B.~P.~Abbott {\it et al.} (Virgo, LIGO Scientific),
  %%Observation of Gravitational Waves from a Binary Black Hole Merger.
  {\it Phys.\ Rev.\ Lett.\ } {\bf 116} (2016), 061102.
  %%CITATION = doi:10.1103/PhysRevLett.116.061102;%%
  %4072 citations counted in INSPIRE as of 31 Jul 2019

%\cite{Abbott:2016nmj}
\bibitem{Abbott:2016nmj}
  B.~P.~Abbott {\it et al.} (Virgo, LIGO Scientific and Virgo Collaborations),
  %%GW151226: Observation of Gravitational Waves from a 22-Solar-Mass Binary Black Hole Coalescence.
  {\it Phys.\ Rev.\ Lett.\ } {\bf 116} (2016), no. 24, 241103.
  %%CITATION = doi:10.1103/PhysRevLett.116.241103;%%
  %1843 citations counted in INSPIRE as of 31 Jul 2019

%\cite{Abbott:2017vtc}
\bibitem{Abbott:2017vtc}
  B.~P.~Abbott {\it et al.} (Virgo, LIGO Scientific),
  %%GW170104: Observation of a 50-Solar-Mass Binary Black Hole Coalescence at Redshift 0.2.
  {\it Phys.\ Rev.\ Lett.\ } {\bf 118} (2017), 221101
  Erratum: [{\it Phys.\ Rev.\ Lett.\ } {\bf 121} (2018), 129901].
  %%CITATION = doi:10.1103/PhysRevLett.118.221101, 10.1103/PhysRevLett.121.129901;%%
  %1279 citations counted in INSPIRE as of 31 Jul 2019

%\cite{TheLIGOScientific:2017qsa}
\bibitem{TheLIGOScientific:2017qsa}
  B.~P.~Abbott {\it et al.} (Virgo, LIGO Scientific),
  %%GW170817: Observation of Gravitational Waves from a Binary Neutron Star Inspiral.
  {\it Phys.\ Rev.\ Lett.\ } {\bf 119} (2017), 161101.
  %%CITATION = doi:10.1103/PhysRevLett.119.161101;%%
  %123 citations counted in INSPIRE as of 14 Nov 2017

%\cite{George:2016hay}
\bibitem{George:2016hay}
  D.~George and E.~A.~Huerta,
  %%Deep Neural Networks to Enable Real-time Multimessenger Astrophysics.
  {\it Phys.\ Rev.\ D} {\bf 97} (2018), 044039.
  %%CITATION = doi:10.1103/PhysRevD.97.044039;%%
  %33 citations counted in INSPIRE as of 31 Jul 2019

%\cite{George:2017pmj}
\bibitem{George:2017pmj}
  D.~George and E.~A.~Huerta,
  %%Deep Learning for Real-time Gravitational Wave Detection and Parameter Estimation: Results with Advanced LIGO Data.
  {\it Phys.\ Lett.\ B} {\bf 778} (2018), 64.
  %%CITATION = doi:10.1016/j.physletb.2017.12.053;%%
  %31 citations counted in INSPIRE as of 31 Jul 2019

%\cite{Gabbard:2017lja}
\bibitem{Gabbard:2017lja}
  H.~Gabbard, M.~Williams, F.~Hayes and C.~Messenger,
  %%Matching matched filtering with deep networks for gravitational-wave astronomy.
  {\it Phys.\ Rev.\ Lett.\ } {\bf 120} (2018), 141103.
  %%CITATION = doi:10.1103/PhysRevLett.120.141103;%%
  %11 citations counted in INSPIRE as of 31 Jul 2019

%\cite{Abbott:2017gyy}
\bibitem{Abbott:2017gyy}
  B.~P.~Abbott {\it et al.} (Virgo, LIGO Scientific),
  %%GW170608: Observation of a 19-solar-mass Binary Black Hole Coalescence.
  {\it Astrophys.\ J.\ } {\bf 851} (2017), L35.
  %%CITATION = doi:10.3847/2041-8213/aa9f0c;%%
  %555 citations counted in INSPIRE as of 04 Aug 2019

%\cite{GBM:2017lvd}
\bibitem{GBM:2017lvd}
  B.~P.~Abbott {\it et al.} (Virgo, LIGO Scientific),
  %%Multi-messenger Observations of a Binary Neutron Star Merger.
  {\it Astrophys.\ J.\ }  {\bf 848} (2017), L12.
  %%CITATION = doi:10.3847/2041-8213/aa91c9;%%
  %535 citations counted in INSPIRE as of 16 Dec 2018

%\cite{Canton:2014ena}
\bibitem{Canton:2014ena}
  T.~Dal Canton {\it et al.},
  %%Implementing a search for aligned-spin neutron star-black hole systems with advanced ground based gravitational wave detectors.
  {\it Phys.\ Rev.\ D} {\bf 90} (2014), 082004.
  %%CITATION = doi:10.1103/PhysRevD.90.082004;%%
  %96 citations counted in INSPIRE as of 04 Aug 2019

%\cite{Harry:2016ijz}
\bibitem{Harry:2016ijz} 
  I.~Harry, S.~Privitera, A.~Bohé and A.~Buonanno,
  %``Searching for Gravitational Waves from Compact Binaries with Precessing Spins,''
  {\it Phys. Rev. D} {\bf 94}, no. 2, (2016) 024012.
  %%CITATION = doi:10.1103/PhysRevD.94.024012;%%
  %50 citations counted in INSPIRE as of 12 Jan 2020

%\cite{Huerta:2016rwp}
\bibitem{Huerta:2016rwp} 
  E.~A.~Huerta {\it et al.},
  %``Complete waveform model for compact binaries on eccentric orbits,''
  {\it Phys.\ Rev.\ D} {\bf 95}, no. 2, (2017) 024038.
  %%CITATION = doi:10.1103/PhysRevD.95.024038;%%
  %51 citations counted in INSPIRE as of 12 Jan 2020

%\cite{Huerta:2019rtg}
\bibitem{Huerta:2019rtg} 
  E.~A.~Huerta {\it et al.},
  %``Enabling real-time multi-messenger astrophysics discoveries with deep learning,''
  {\it Nature Reviews Physics} {\bf 1}, (2019) 600-608.
  %%CITATION = doi:10.1038/s42254-019-0097-4;%%
  %1 citations counted in INSPIRE as of 12 Jan 2020

%\cite{Metzger:2011bv}
\bibitem{Metzger:2011bv}
  B.~D.~Metzger and E.~Berger,
  %%What is the Most Promising Electromagnetic Counterpart of a Neutron Star Binary Merger?
  {\it Astrophys.\ J.\ } {\bf 746} (2012), 48.
  %%CITATION = doi:10.1088/0004-637X/746/1/48;%%
  %368 citations counted in INSPIRE as of 04 Aug 2019

\bibitem{DL} Y.~LeCun, Y.~Bengio and G.~Hinton, Deep Learning. {\it Nature}, {\bf 521} (2015), 436..

\bibitem{DLVis} K.~He, X.~Zhang, S.~Ren and J.~Sun, Deep Residual Learning for Image Recognition.
{\it The IEEE Conference on Computer Vision and Pattern Recognition (CVPR)}, {\bf 770} (2016).

\bibitem{DL_NLP} T.~Young, D.~Hazarika, S.~Poria and E.~Cambria, Recent Trends in Deep Learning Based Natural Language Processing.
{\it IEEE Computational Intelligence Magazine} {\bf 13} (2018), 55-75.
%\cite{Wei:2019zlc}

\bibitem{Wei:2019zlc} 
  W.~Wei and E.~A.~Huerta,
  %``Gravitational Wave Denoising of Binary Black Hole Mergers with Deep Learning,''
  {\it Phys. Lett. B} {\bf 800} (2020), 135081.
  %%CITATION = doi:10.1016/j.physletb.2019.135081;%%
  %11 citations counted in INSPIRE as of 12 Jan 2020

\bibitem{DL_Book} I.~Goodfellow, Y.~Bengio, and A.~Courville, Deep Learning (MIT Press, Cambridge, MA, 2016).

\bibitem{LALSuite} LIGO Scientific Collaboration, LIGO Algorithm Library -- LALSuite (GPL 2018).

%\cite{Abbott:2018wiz}
\bibitem{Abbott:2018wiz}
  B.~P.~Abbott {\it et al.} (Virgo and LIGO Scientific),
  %%Properties of the binary neutron star merger GW170817.
  {\it Phys.\ Rev.\ X } {\bf 9} (2019), 011001.
  %%CITATION = doi:10.1103/PhysRevX.9.011001;%%
  %203 citations counted in INSPIRE as of 05 Aug 2019

\bibitem{Akmal:1998cf}A.~Akmal, V.~R.~Pandharipande, D.~G.~Ravenhall,
%%Equation of state of nucleon matter and neutron star structure.
{\it Phys. Rev. C} {\bf 58} (1998), 1804.

%\cite{Hinderer:2009ca}
\bibitem{Hinderer:2009ca}
  T.~Hinderer, B.~D.~Lackey, R.~N.~Lang and J.~S.~Read,
  %%Tidal deformability of neutron stars with realistic equations of state and their gravitational wave signatures in binary inspiral.
  {\it Phys.\ Rev.\ D} {\bf 81} (2010), 123016.
  %%CITATION = doi:10.1103/PhysRevD.81.123016;%%
  %150 citations counted in INSPIRE as of 06 Jul 2017

%\cite{Krastev:2018wtx}
\bibitem{Krastev:2018wtx}
  P.~G.~Krastev and B.~A.~Li,
  %%Imprints of the nuclear symmetry energy on the tidal deformability of neutron stars.
  {\it J.\ Phys.\ G} {\bf 46} (2019), 074001.
  %%CITATION = doi:10.1088/1361-6471/ab1a7a;%%
  %22 citations counted in INSPIRE as of 06 Aug 2019

%\cite{Purrer:2015tud}
\bibitem{Purrer:2015tud}
  M.~Pürrer,
  %%Frequency domain reduced order model of aligned-spin effective-one-body waveforms with generic mass-ratios and spins.
  {\it Phys.\ Rev.\ D} {\bf 93} (2016), 064041.
  %%CITATION = doi:10.1103/PhysRevD.93.064041;%%
  %78 citations counted in INSPIRE as of 05 Aug 2019

\bibitem{LIGO-LRR-2016} B.~P.~Abbott {\it et al.} (Virgo and LIGO Scientific),
%%Prospects for Observing and Localizing Gravitational-Wave Transients with Advanced LIGO and Advanced Virgo.
{\it Living Rev. Relativ.}, {\bf 19} (2016), 1.

%\cite{Shen:2019vep}
\bibitem{Shen:2019vep}
  H.~Shen, E.~A.~Huerta and Z.~Zhao,
  Deep Learning at Scale for Gravitational Wave Parameter Estimation of Binary Black Hole Mergers.
  arXiv:1903.01998.
  %%CITATION = ARXIV:1903.01998;%%

\bibitem{CNN} Y.~Lecun, L.~Bottou, Y.~Bengio and P.~Haffner,
  %%Gradient-based learning applied to document recognition. 
  {\it Proc. IEEE} {\bf 86} (1998), 2278-2324.

\bibitem{BP} Y.~Hirose, K.~Yamashita and S.~Hijiya, 
%%Back-propagation algorithm which varies the number of hidden units.
{\it Neural Networks} {\bf 4} (1991), 61-66.

\bibitem{TF} Martín Abadi, Ashish Agarwal, Paul Barham, Eugene Brevdo,
Zhifeng Chen, Craig Citro, Greg S. Corrado, Andy Davis,
Jeffrey Dean, Matthieu Devin, Sanjay Ghemawat, Ian Goodfellow,
Andrew Harp, Geoffrey Irving, Michael Isard, Rafal Jozefowicz, Yangqing Jia,
Lukasz Kaiser, Manjunath Kudlur, Josh Levenberg, Dan Mané, Mike Schuster,
Rajat Monga, Sherry Moore, Derek Murray, Chris Olah, Jonathon Shlens,
Benoit Steiner, Ilya Sutskever, Kunal Talwar, Paul Tucker,
Vincent Vanhoucke, Vijay Vasudevan, Fernanda Viégas,
Oriol Vinyals, Pete Warden, Martin Wattenberg, Martin Wicke,
Yuan Yu, and Xiaoqiang Zheng.
TensorFlow: Large-scale machine learning on heterogeneous systems,
2015. Software available from tensorflow.org.

\bibitem{ADAM} D. P. Kingma and J. Ba, Adam: A Method for Stochastic Optimization, arXiv:1412.6980.

\bibitem{ADAM2} S. J. Reddi, S. Kale, and S. Kumar, On the convergence of Adam and beyond, arXiv:1904.09237.

\bibitem{Gebhard:2019ldz} 
 T.~D.~Gebhard, N.~Kilbertus, I.~Harry and B.~Schölkopf,
 %``Convolutional neural networks: a magic bullet for gravitational-wave detection?,''
 {\it Phys. Rev. D} {\bf 100} (2019) 063015.
  %%CITATION = doi:10.1103/PhysRevD.100.063015;%%

\end{thebibliography}
\end{document}